\documentclass[sigplan]{acmart}

\setcopyright{acmlicensed}

\copyrightyear{2024}
\acmYear{2024}
\setcopyright{acmlicensed}\acmConference[PaPoC '24]{The 11th Workshop on Principles and Practice of Consistency for Distributed Data}{April 22, 2024}{Athens, Greece}
\acmBooktitle{The 11th Workshop on Principles and Practice of Consistency for Distributed Data (PaPoC '24), April 22, 2024, Athens, Greece}
\acmDOI{10.1145/3642976.3653036}
\acmISBN{979-8-4007-0544-1/24/04}

\usepackage{csquotes}

\newcommand{\todo}[2][]{}

\newcommand{\transaction}[2]{\begin{color}{#1}#2\end{color}}

\usepackage{listings}
\usepackage{xcolor}

\definecolor{t0}{RGB}{0,51,102}   % Navy Blue
\definecolor{t1}{RGB}{204,51,51}  % Dark Red
\definecolor{t2}{RGB}{0,102,51}   % Dark Green
\definecolor{t3}{RGB}{255,140,0}  % Dark Orange
\definecolor{t4}{RGB}{102,0,102}  % Dark Purple

\definecolor{gray}{rgb}{0.5,0.5,0.5}
\lstdefinestyle{scitzenCodestyle}{
  frame=none,
  showstringspaces=false,
  basicstyle={\ttfamily},
  numbers=left,
  xleftmargin=2.3em,
  numberstyle=\fontsize{7}{9}\selectfont\color{gray}\bfseries,
  breaklines=true,
  breakatwhitespace=true,
  tabsize=2,
  escapeinside={(*@}{@*)},
  numberblanklines=true,
  firstnumber=last,
  literate={"}{\textquotedbl}1,
}
\usepackage{subcaption}
\usepackage{graphicx}  %% Graphics
    \setcitestyle{numbers,square,sort&compress}
\usepackage[defblank]{paralist}  %% inline lists
\usepackage{parskip} \parskip=1ex plus 5pt minus 1pt \parindent=4ex %% indent paragraphs
\usepackage{xcolor}  %% Use of color
\usepackage{xspace}  %% Macros with automatic spacing
\usepackage{algpseudocode}
\usepackage{tikz}
\usetikzlibrary{positioning}

\usepackage{hyperref}  %% hyperlinks in PDF (must be last package loaded)

\usepackage{ourmacros}
\usepackage{types}

\usepackage{mathpartir} % Formatting inference rules

\begin{document}

\title{Models for Storage in Database Backends} 
\subtitle {A Rigorous Approach for Formally-Correct Designs}

\author{Edgard Schiebelbein}
 % \orcid{}
 \affiliation{
   \institution{University of Kaiserslautern-Landau}
   %\city{Kaiserslautern}
   \country{Kaiserslautern, Germany}
   }
 \email{e_schiebel19@cs.uni-kl.de}

\author{Saalik Hatia}
 \orcid{0000-0003-2747-118X}
 \affiliation{
   \institution{Sorbonne-Université (LIP6)}
   %\city{Paris}
   \country{Paris, France}
   }
 \email{saalik.hatia@lip6.fr}

\author{Annette Bieniusa}
 \orcid{0000-0002-1654-6118}
 \affiliation{
   \institution{University of Kaiserslautern-Landau}
   %\city{Kaiserslautern}
   \country{Kaiserslautern, Germany}
   }
 \email{bieniusa@cs.uni-kl.de}

\author{Gustavo Petri}
 \authornote{This work was conducted while Gustavo Petri was at ARM Ltd.}
 \orcid{0000-0003-3289-4574}
 \affiliation{
   \institution{Amazon Web Services}
   %\city{Cambridge}
   \country{Cambridge, United Kingdom}
 }
% \email{Gustavo.Petri@arm.com}

\author{Carla Ferreira}
 \orcid{0000-0003-3680-7634}
 \affiliation{
   \institution{Universidade NOVA de Lisboa}
   %\city{Lisbon}
   \country{Lisbon, Portugal}
   }
 \email{carla.ferreira@fct.unl.pt}

\author{Marc Shapiro}
 \orcid{0000-0002-8953-9322}
 \affiliation{
   \institution{Sorbonne-Université (LIP6) \& Inria}
   %\city{Paris}
   \country{Paris, France}
 }
 % \affiliation{
 %   \institution{Inria}
 %   \city{Paris}
 \email{marc.shapiro@acm.org}

\renewcommand{\shortauthors}{Schiebelbein \emph{et al.}}


%% Local Variables:
%% mode: latex
%% coding: utf-8
%% ispell-local-dictionary: "en_GB"
%% mode: flyspell
%% TeX-master: "paper" 
%% my-latex-bibfile: "~/svnbackup/bib/shapiro-bib-ext.bib"
%% End:

\begin{abstract}
  This paper describes ongoing work on developing a formal specification of a database backend. 
  % in order to support sophisticated features such as caching, write-ahead
  % logging, checkpointing, or journal truncation.
  % Our insight is such features can be described and formalised by composing
  % fundamental feature variants.
We present the formalisation of the expected behaviour of a basic transactional system that calls into a simple store API, and
 instantiate in two semantic models. % of a store.
The first one is a map-based, classical versioned key-value store; the second one, journal-based, appends individual transaction effects to a journal. 
We formalise a significant part of the specification in the Coq proof assistant. 
%We aim to formally show that both semantic variants are behaviourally equivalent to the store API specification, and are thus interchangeable.
This work will form the basis for a formalisation of a full-fledged backend store with features such as caching or write-ahead logging, as variations on maps and journals.
\end{abstract}

\ccsdesc[300]{Theory of computation~Operational semantics}
\ccsdesc[500]{Information systems~Information storage systems}

\keywords{formal methods, verification, key-value store}


\maketitle

\section{Introduction}
A database system manages a collection of digital data.
An essential component is the \emph{backend}, which is in charge of recording the data into some memory or \emph{store}.
Although conceptually simple at a high level, actual backends are
complex, due to the demands for fast response, high volume, limited
footprint, concurrency, distribution, and reliability.
For instance, the open-source RocksDB comprises 350+\,kLOC and Redis is
approximately 200\,kLOC \cite{rocksdbgithub,redisgithub}.
Any such complex software has bugs; and database backend bugs are critical,
possibly violating data integrity or security~\cite{rocksdbbug, jepsen}.

\todo[Marc]{Cite some episodes}

Formal methods have the potential to avoid such bugs, but,
given the complexity of a modern backend, fully specifying all the
moving pieces is a daunting task.
% Furthermore, formally-verified development is often perceived by
% developers as at odds with performance.
% To our knowledge, formally-verified development of a full-fledged storage
% backend has not been attempted before.

This paper reports on an incremental approach to the rigorous and modular
development of such a backend towards an implementation.
To this end, we formalise the semantics of atomic transactions above a versioned key-value store; this high-level specification helps to reason about correctness, both informally and formally with the Coq proof tool.
% In the style of MVCC (multi-value concurrency control), a transaction
% reads from a causally-consistent snapshot.
% It terminates by either committing atomically, or by aborting without
% modifying the store.
Although this paper focuses on a highly-available transaction model
(convergent causal consistency or TCC+, a variant of PSI
\cite{rep:syn:1661}), our results generalise to stronger
models such as SI or strong serialisability.
The transaction model appeals to a store's specialised book-keeping
operations (called \DoBegin{} \DoUpdate{}, \DoCommit{}, and \Lookup{}),
implicity assuming infinite memory and no failures.
%To bound a store's memory footprint, we restrict its domain.

Next, we instantiate these semantics with two models of the store.
The first variant is a classical map-based, versioned key-value store.
As a transaction executes, it \emph{eagerly} computes new versions, which it copies into a map upon commit, labelled with the transaction's commit timestamp; reading a key searches the map for the most recent corresponding version.
We plan to mechanise the proof that the map-based model satisfies the transactional specification, i.e., that in any reachable state, a call to \Read{} returns the value expected by the semantics
%noted {\SafeInv{\aTaggedEffectSet}{\aKey}{\aTS}} 
for any key and timestamp pair.
%We translate the formal specification verbatim into Java, forsaking
%optimisations, and test empirically that this implementation satisfies
%the specification.

Our second variant uses a journal (or log).
A transaction appends individual effects to the journal, tagged with the transaction identifier; committing appends a commit record, sealing it with its commit timestamp.
Reading from the journal applies all the relevant effects previously recorded in the journal.
Again, we plan to prove mechanically that the journal-based store satisfies the abstract specification.
%We transcribe the specification to a Java crash-tolerant journal store,
%and show empirically that it satisfies the specification.

% Reading the specification as a kind of pseudocode, we implement it
% verbatim, without optimisation.
% More specifically, we implement a map-based and a journal-based store,
% both in memory and on disk.
% Experimentally, as expected, the former has high performance but is
% limited by memory size; the on-disk map is persistent (but not crash
% tolerant) and IO-bound; the on-disk journal has high throughput but slow
% response.

In this paper, we summarize our work-in-progress on the formal models for the journal- and map-based store.
We define a common interface and show how these stores can be employed in a transactional storage system.
Further, we sketch their implementation and reasoning about their correctness in Coq.

The models presented here lack fault-tolerance and essential features such as sharding, ca\-ching, write-ahead logging, etc., which are required for state-of-the-art performance.
Our hypothesis for future work is that such features can be described and
implemented by \emph{composing} instances of these basic variants.
%In particular, we show how to compose a write-ahead log and to bound
%storage footprint.
% In particular, we show how to build layered storage (as in LSM Tree) and
% to bound storage footprint.
% We further believe the same approach can justify sharding,
% and geo-distribution.
%The formal rules for correct dynamic composition are particularly
%simple.

% This short paper contributes a formal model of a concurrent,
% transactional backend store, and its map- and journal-based variants.
% The proofs of correctness and of behavioural equivalence, as well as a
% concrete implementation inspired by the specification, are in progress.
  
% The paper proceeds as follows.
% After this introduction, Section~\ref{sec:system-model} formalises the
% system model and semantics.
% Section~\ref{sec:transactions} presents our system model, APIs,
% and basic concepts.
% Section~\ref{sec:transactions} describes transactions, and
% Section~\ref{sec:stores} details the map and journal stores.
% Our work so far with Coq is described in Section~\ref{sec:coq}.
% We conclude with a  brief comparison with the state of the art and a
% discussion of future work, in Section~\ref{sec:discussion-outlook}.
% Appendix~\ref{app:transaction-semantics} presents a formal semantics
% of transactions for interested readers.

\begin{table*}[tp]
\begin{minipage}{0.9\hsize}\centering \small
\begin{tabular}[t]{@{}ccl@{}}
  \aKey, \aTS, \aValue             & $\in \TKey, \in \TTimestamp, \in \TValue$                               & key, timestamp, value                   \\
  \aEffect[]                       & $\in \TEffect = \TValue_{\bottom} \rightarrow \TValue $                  & effect                                 \\
  \bottom                          & $\notin \TEffect$
                                                                                                             & absence of effect \todo[AB]{overloaded to empty{\slash}undefined}                     \\
  $\apply$                   & $\in \TEffect_{\bottom} \times \TEffect_{\bottom} \rightarrow \TEffect_{\bottom}$ & effect composition         \\
  \hline
  \aTxn{}                           & $\in \TTxn$                                                             & Transaction \\
  \aTxnID                           & $\in \TTxnID$                                                           & transaction identifier                \\
  \aST                              & $\in \TTimestamp$                                                       & snapshot timestamp (of transaction) \\
  \aReadSet                         & $\in \mathcal{P}(\TKey)$                                                & read set, keys read in transaction \\
  \aWriteSet                        & $\in \mathcal{P}(\TKey)$                                                & write{\slash}dirty set, keys modified in txn \\
  \aEffectBuf                       & $\in \TEffectBuf = \TKey \rightarrow \TEffect_{\bottom}$                 & effect buffer (of transaction)          \\
  \aCT                              & $\in \TTimestamp$                                                       & commit timestamp (of transaction)     \\
  \TxnDescriptorDefault             & $\in \TTxnDescriptor$
                                     % typedef not necessary: duplicates the previous lines
                                     %   $\in \TTxnDescriptor = \TTxnID \times \TTimestamp~\times$ & \\
                                     % & $\mathcal{P}(\TKey) \times \mathcal{P}(\TKey)
                                     %   \times \TEffectBuf \times \TTimestamp$                                  
                                                                                                              & transaction descriptor \\
  \hline

  \abortedSet, \committedSet, \runningSet    & $\subseteq \TTxnDescriptor$          & aborted, committed, running transactions \\

  % \abortedSet                          & $\subseteq \TTxnDescriptor$                                          & aborted transactions \\
  % \committedSet                        & $\subseteq \TTxnDescriptor$                                        & committed transactions \\
  % \runningSet                          & $\subseteq \TTxnDescriptor$                                          & running transactions \\
  % ~                                 &                                                                         &                                       \\
\end{tabular}
\end{minipage}
\caption{Overview of notation.}
  \label{tbl:notation}
\end{table*}

\begin{figure}[tp]
\begin{minipage}{0.9\hsize}\centering \small
\[
  \begin{array}{rcl}
    \Empty{}   & : & \TStore\\
    \DoBegin   & : & \TStore
                      \times \TTxnID
                      \times \TTimestamp 
                      \rightarrow \TStore \\
    \Lookup{}   & : & \TStore
                      \times \TKey
                      \times \TTimestamp 
                      \rightarrow \TEffect_{\bottom}\\
    \DoUpdate{} & : & \TStore
                      \times \TTxnID
                      \times \TKey
% \aST is redundant   \times \TTimestamp
                      \times \TEffect 
                      \rightarrow \TStore \\
    \DoCommit{} & : & \TStore
                      \times \TTxnID
                      \times \TTimestamp
                      \times \mathcal{P}(\TKey) \\
                &   & \times \mathcal{P}(\TKey)
                      \times \TEffectBuf
                      \times \TTimestamp
                      \rightarrow \TStore\\
  \end{array}
\]
\end{minipage}
\vspace{-5pt}
\caption{Store interface.}
  \label{fig:store-interface}
\end{figure}

\section{System Model and Terminology}
\label{sec:system-model}

Next, we present an informal, high-level overview of the system model
and terminology.
% In the following sections, we will formalise these concepts precisely, discuss
% their properties, and present our journey of developing a formal specification to implementation. 
Table~\ref{tbl:notation} overviews our notation.

\paragraph{Stores and transactions}
At the core of the model is an abstract mutable shared memory, called a \emph{store}, 
$\aStore \in \TStore$.
% A store associates keys, $\aKey \in \TKey$,  with values, $\aValue \in \TValue$.
A store follows the common API shown in
Figure~\ref{fig:store-interface}.
Method \Lookup{} returns the value that store \aStore{} associates with key \aKey{} at time \aTS{}; an absent mapping returns \bottom{} (i.e., initially, every key maps to \bottom).
Update method \DoUpdate{}  applies a new \emph{effect} \aEffect{}
to that key's entry in the store, in order to update the value (see below).
Successfully invoking \DoCommit{} makes the updates of the
current transaction visible with a commit timestamp noted {\aCT}.
% In general, all operations other than  \Lookup{} return a (potentially) modified store.

Updating a key \aKey{} under timestamp \aTS{} creates a new version
mapped at index $(\aKey, \aTS{})$.
A mapping is write-once, and remains valid until the next mapping, if any.
For example, suppose store \aStore{} updates a version of key \aKey{} at
time $\aTS = 100$  with an assignment of 27.%
\footnote{
  Assuming integer timestamps, for the sake of example.
}
Then, \LookupInv{\aStore{}}{\aKey{}}{101} should return 27.
If there are no other versions between 100 and 110,
\LookupInv{\aStore{}}{\aKey{}}{111} should also return 27.
If the next mapping is at timestamp 120, to \Incr[]{10}, then 
\LookupInv{\aStore{}}{\aKey{}}{121} should return 37.

A client (left unspecified by our model) accesses the store in the
context of a \emph{transaction}, a sequence of begin, lookup, update and
commit{\slash}abort actions.
A transaction reads from
a consistent snapshot of the store, and makes its effects visible in the store by committing the transaction atomically 
(all-or-nothing).
We defer a more detailed discussion of transactions to Section~\ref{sec:transactions}.

\paragraph{Keys, values and timestamps}
Keys, values, and timestamps  are opaque types.
% , denoted by the sets
% $\TKey$, $\TValue$, and $\TTimestamp$.
Keys compare for equality only.
Timestamps are partially ordered by $\leq$;
% ; i.e. $\le$ is reflexive, antisymmetric, and transitive.
we say timestamps are \emph{concurrent}, if they are not ordered, i.e., 
$\aTS[1] \concur \aTS[2] \equaldef \aTS[1] \not\leq
\aTS[2] \land \aTS[2] \not\leq \aTS[1]$.
Note that we do not assume a global clock.
% Two timestamps have a least upper bound $\sup$, and a greatest lower
% bound $\inf$.
% We generalise $\inf$ and $\sup$ to a set in the standard fashion; as a
% shorthand, $\sup_{\aTxnSet{}}(x) \equaldef \sup(
% \Pi_{x}(\aTxnSet{}) )$ is the least-upper-bound of the $x$
% dimension in the set of tuples $\aTxnSet{}$.
% We note $\aTS[1] \gepar \aTS[2] \equaldef \aTS[1] \not<
% \aTS[2]$, read ``greater, equal, or concurrent.''
% The notations $\inf$ and $\lepar$ are defined symmetrically.
% For full generality, the timestamp type may be represented as a vector
% of integers, with the classical definitions for $\le$
% \cite{alg:rep:738, alg:rep:738bis}.
% In this case, we define the least upper bound, or supremum, $\sup$ as follows:
%   $\forall i: \sup(\aTS[1],\aTS[2])[i] =
%   \sup(\aTS[1][i],\aTS[2][i])$; and similarly for $\inf$.
% Our model also works for total order, using scalar timestamps.

\todo[TODO: Replication]{%
  Any participant [def???] in the system may have its local clock,
  noted \now{}, such that only events timestamped $\le \now$ are visible
  to it; if it attempts an operation with a larger timestamp, the operation
  will not return until \now{} increases sufficiently.
}
% A \emph{clock} is an object that returns unique and monotonically-increasing timestamps.

An \emph{ordered timestamp pair} (OTSP) is of the form $(d,v) \in \TTimestamp \times \TTimestamp$, where $d \le v$, called dependence and version respectively.
We define a strict partial order relation $\visOTSP$ over OTSPs, as follows:
$(d_{1}, \aVT[1]) \visOTSP (d_{2}, \aVT[2])
  \equaldef
  \aVT[1] < d_{2}
$.
Two OTSPs are concurrent if they cannot be ordered by $\visOTSP$:
\begin{small}
\begin{align*}
(d_{1}, \aVT[1]) &\concurOTSP (d_{2}, \aVT[2])  \\
  & \equaldef  (d_{1}, \aVT[1]) \not\visOTSP (d_{2}, \aVT[2]) \land (d_{2}, \aVT[2]) \not\visOTSP (d_{1}, \aVT[1]) \\
  &\equiv (d_{2} \le \aVT[1] \lor  d_{2} \concur \aVT[1]) \land (d_{1} \le \aVT[2] \lor  d_{1} \concur \aVT[2])
\end{align*}
\end{small}

\paragraph{Effects}
Classically, an update simply assigns a new value to the key, as in
$\aKey\aAssign[27]$. Such an assignment creates a new version of \aKey{} with value
27. 
% We assume that all versions of a same key have the same data type.

Many recent stores \cite{rep:lan:1869, syn:rep:sh143, app:rep:1800}
support a more general concept of update, which we call \emph{effect}.
%An {effect} $\aEffect \in \TEffect$ is a function from value to
%value.
Applying effect \aEffect{} to a current value \aValue{} computes a new value $\aEffect(\aValue)$.
For instance, $\delta_{incr10}$ is the effect that adds 10 to some value.
An assignment is thus a constant effect; for example, $\delta_{\text{assign}_{27}}(\aValue{})$
yields 27 whatever the current value \aValue{}.
% %
% \footnote{%
% %
%   Technically, our formalism does not support assigning \bottom{}, i.e.,
%   a key that has been assigned a value cannot be deleted from the store thereafter.
%   A more general handling would be at the expense of a more complex formalism which we will skip for now.
% }
%

If a sequence of updates with effects $\aEffect, \aEffect', \ldots{}, \aEffect''$ has been applied to key \aKey{}, then a store is expected to return value $(\aEffect \apply \aEffect' \apply \ldots{} \apply \aEffect'')(\bottom)$ when queried.%
\footnote{
  Operator $\apply$ is pronounced ``apply;'' it is equivalent to
  classical functional composition $\circ$, but associates left to right
  for convenience: $\aEffect \apply \aEffect' = \aEffect' \circ \aEffect$.
  % , defined as $\forall \aValue : \aValue \apply \aEffect \apply \aEffect' = (
  % \aEffect' \circ \aEffect )(\aValue) \equaldef
  % \aEffect'(\aEffect(\aValue))$.
  Note that \apply{} is associative.
}
We say a sequence of effects is \emph{proper} if it starts with an assignment and therefore evaluates to a value, or $\bottom$, when applied to $\bottom$; i.e., it does not depend on any preceding effects.
% In particular, a sequence that starts with an assignment is proper.
% The application of effects of a proper sequence is well-defined, since
% we assume that entries in a store are initially undefined ($\bottom$).
% In what follows,
We assume that every history forms a proper sequence.

An assignment masks any previous effects to the same key: $\forall \aEffect, (\aEffect \apply \aEffect[{\text{assign}}]) = \aEffect[{\text{assign}}]$; therefore effects that precede the last assignment in a sequence can be safely ignored.
Conversely, any proper sequence is equivalent to a single assignment.
For instance, $\aEffect[{\text{assign}_{27}}] \apply \aEffect[{\text{incr10}}] = \aEffect[{\text{assign}_{37}}]$.
This justifies checkpointing a proper sequence into a single
assignment.
Figure~\ref{fig:apply} summarizes the rules of effect composition.

%\todo[Marc]{Define \emph{relevant} subsequence: omits effects that precede assignment.}
  
\begin{figure}[tp]
  \begin{minipage}{1.0\linewidth} \centering \small
    \[
      \begin{array}[c]{cl}
        (\bottom \apply \aEffect) = (\aEffect \apply \bottom) = \aEffect   & \text{(non-effect)} \\
        (\aEffect \apply \aEffect') \apply \aEffect''
            = \aEffect \apply (\aEffect' \apply \aEffect'')                & \text{(associativity)} \\
        (\lambda v.c) \apply \aEffect = \lambda v.\aEffect(c)              & \text{(compacting a proper sequence)} \\
      \end{array}
    \]
  \end{minipage}
  \vspace{-5pt}
  \caption{Effect composition.}
  \label{fig:apply}
\end{figure}

\paragraph{Visibility and concurrent effects}

%\todo[AB]{Problem: Semantics of empty effect: For merge, the empty effect could be identity; the problem is what happens if we apply an empty effect on bottom value}
%Effects are identified by their \emph{timestamp} $\aTS$.

Effects are ordered by the \emph{visibility} relation $\aEffect \vis
\aEffect'$ (read ``\aEffect is visible to \aEffect' ''), defined as
follows:
\begin{compactitem}
\item % [(Internal visibility INT)]
  $\aEffect \vis \aEffect'$ if both belong to the same transaction,
  and $\aEffect$ is before $\aEffect'$.
\item % [(External visibility EXT)]
  $\aEffect \vis \aEffect'$ if they belong to different transactions,
  $\aTxn{}$ and $\aTxn{}'$ respectively,
  % where $\aTxn{}'$ can read from  $\aTxn{}$; i.e.,
  where $\aTxn{}$ has committed, and \aTxn{} is before $\aTxn'$ in OTSP
  order, i.e., $\aTxn{}.\aCT < \aTxn{}'.\aST$.
\end{compactitem}
Visibility is a strict partial order.
Two effects are concurrent if they are not mutually ordered by visibility. 

%if, either their transactions' commit timestamps are concurrent, or if
%their timestamp ranges overlap.
% \[  \aTxn{} \concur \aTxn{}' \equaldef (\aTxn{}.\aST, \aTxn{}.\aCT{}) \concurTSP (\aTxn{}'.\aST, \aTxn{}'.\aCT{})\]
% %
%If timestamps are totally ordered in the system, transactions might still be concurrent (for instance, under Snapshot
% Isolation).
% Note also that the concurrency relation is undefined for an uncommitted transaction.

Some data types support concurrent effects thanks to a \Merge{} operator
on effects.
% Classical sequential data types generally disallow concurrent updates,
% leaving \Merge{} undefined.
To ensure convergence, the \Merge{} operator is required to be
commutative, associative, and idempotent (CAI) \cite{syn:rep:sh143}.

In the presence of concurrent effects, the value expected of key \aKey
is results from applying, from the initial \bottom{}, the visible
effects related to \aKey{}, in visibility order, while \Merge{}ing
concurrent effects.

\paragraph{Concurrent data types}

As an aside, note that classical sequential data types generally
disallow concurrent updates, leaving \Merge{} undefined.
These data types require a strong consistency model, %one 
where updates
occur in some serial order.

Data types that merge concurrent effects do exist \cite{syn:rep:sh143}.
There are also data types designed with non-assignment effects \cite{app:rep:1816}.
% ; for
% instance \citet{app:rep:1816} introduce the notion of $\delta$-mutations
% and $\delta$-groups to represent composed and merged effects.

To provide the CAI properties, the implementation of an effect typically needs to carry metadata, e.g., to provide idempotence or determine causal relationships between updates.
For example, the classical last-writer-wins approach supports
concurrency by merging concurrent assignments under some
deterministic total order (e.g., timestamp order), and retaining only
the one with the highest timestamp.
Another example is a counter supporting concurrent increment and decrement effects, which uses a vector of sets of effects, with one entry per (concurrent) client \cite{syn:rep:sh143}.
This representation ensures that a given increment or decrement is applied only once.

\section{Semantics of Transactions}
\label{sec:transactions}

A transaction $\aTxn \in \TTxn$ is a sequence of effects.
We associate to a transaction its \emph{transaction descriptor}
\TxnDescriptorDefault{}.
It reads from a snapshot timestamped by \emph{snapshot timestamp} \aST{}.
% and stages its computations in an \emph{effect buffer} \aEffectBuf{}.
Its \emph{write buffer} \aWriteSet{} lists the keys that it
\emph{dirtied}, i.e., modified.
It may commit with a \emph{commit timestamp} noted \aCT.

However, the effects of a running transaction are not visible from outside (\emph{isolation}).
Within a transaction, an effect \emph{visible} to another one that executes after it (the ``read-your-own-writes'' property \cite{rep:syn:1481}).
The semantics formalise this by staging effects to an \emph{effect
  buffer} \aEffectBuf{}.
% %
% \footnote{
% %
% By associativity, the effect buffer may compress a sequence of effects
% $\aEffect[\aKey]^{1}; \aEffect[\aKey]^{2}; \ldots{};
% \aEffect[\aKey]^{n-1}; \aEffect[\aKey]^{n}$ to key \aKey{}, into a single effect
% $
% (\aEffect[\aKey]^{1}
% \apply
% \aEffect[\aKey]^{2}
% \apply
% \ldots{}
% \aEffect[\aKey]^{n-1}
% \apply \aEffect[\aKey]^{n}).
% $
% As keys are independent, the effects for different keys may occur in any
% order.
% }
% %
A transaction terminates in an all-or-nothing manner, by either an \emph{abort} that discards its effect buffer, or by a \emph{commit} that makes all its effects visible to later transactions at once.
\emph{Atomicity} is formalised by assigning the same, unique,
\emph{commit timestamp} %(noted \aCT{}) 
to all its effects.
The commit timestamp of a running or aborted transaction is irrelevant
and can be arbitrary, marked by $\_$.

Every first read of some key comes from a same \emph{snapshot},
identified by its \emph{snapshot timestamp} noted \aST{}.
Transactions that are \emph{visible} in the snapshot are those that
committed strictly before its snapshot timestamp.

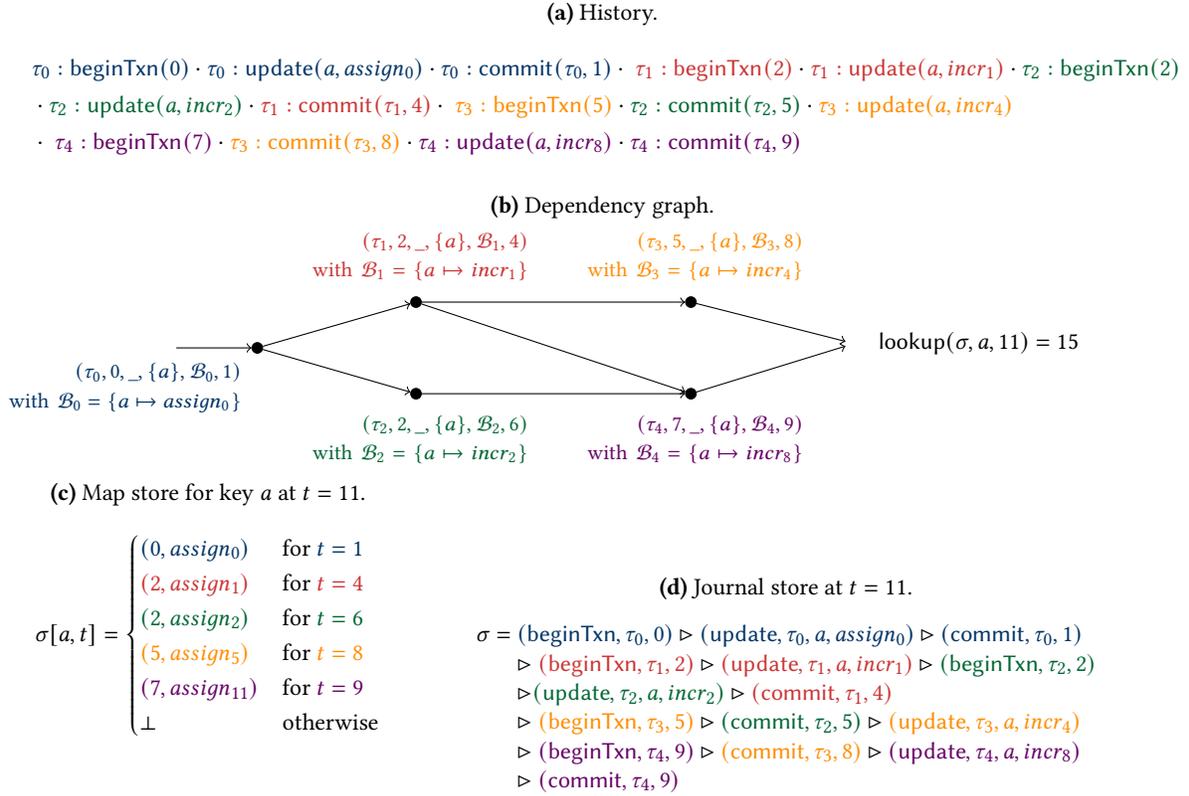
\begin{figure*}[t]
    \begin{minipage}{0.9\hsize}\centering \small

    \begin{subfigure}{1\textwidth}
    \caption{History.}
    \vspace{-5pt}
    \label{fig:ex-history}
    
%     \begin{align*}
%    & \ \aTxnID[0]: \BeginTxnInv{0} \cdot \aTxnID[0]: \UpdateInv{}{x}{assign_0} \cdot  \aTxnID[0]: \CommitTxnInv{\aTxnID[0]}{1} \cdot\ \aTxnID[1] :\BeginTxnInv{2} \cdot \aTxnID[1] : \UpdateInv{}{x}{incr_1} \cdot \aTxnID[2] : \BeginTxnInv{2} \\
%    & \ \cdot \aTxnID[2] :\UpdateInv{}{x}{assign_2} \cdot \aTxnID[1] : \CommitTxnInv{\aTxnID[1]}{4} \cdot\ \aTxnID[3] :\BeginTxnInv{5} \cdot \aTxnID[2] : \CommitTxnInv{\aTxnID[2]}{5} \cdot \aTxnID[3] :\UpdateInv{}{x}{incr_4} \\
%    & \ \cdot\ \aTxnID[4] :\BeginTxnInv{9} \cdot \aTxnID[3] :\CommitTxnInv{\aTxnID[3]}{8} \cdot \UpdateInv{}{x}{incr_8} \cdot \CommitTxnInv{\aTxnID[4]}{9} \\
%    \end{align*}

    \begin{align*}
    & \ \transaction{t0}{\aTxnID[0]: \BeginTxnInv{0}} \cdot 
          \transaction{t0}{\aTxnID[0]: \UpdateInv{}{a}{assign_0}} \cdot  
          \transaction{t0}{\aTxnID[0]: \CommitTxnInv{\aTxnID[0]}{1}} \cdot\ 
          \transaction{t1}{\aTxnID[1] :\BeginTxnInv{2}} \cdot 
          \transaction{t1}{\aTxnID[1] : \UpdateInv{}{a}{incr_1}} \cdot 
          \transaction{t2}{\aTxnID[2] : \BeginTxnInv{2}} \\
    & \ \cdot \transaction{t2}{\aTxnID[2] :\UpdateInv{}{a}{incr_2}} \cdot 
    	 \transaction{t1}{\aTxnID[1] : \CommitTxnInv{\aTxnID[1]}{4}} \cdot\ 
	 \transaction{t3}{\aTxnID[3] :\BeginTxnInv{5}} \cdot 
	 \transaction{t2}{\aTxnID[2] : \CommitTxnInv{\aTxnID[2]}{5}} \cdot 
	 \transaction{t3}{\aTxnID[3] :\UpdateInv{}{a}{incr_4}} \\
    & \ \cdot\ \transaction{t4}{\aTxnID[4] :\BeginTxnInv{7}} \cdot 
    	 \transaction{t3}{\aTxnID[3] :\CommitTxnInv{\aTxnID[3]}{8}} \cdot 
	 \transaction{t4}{\aTxnID[4] :\UpdateInv{}{a}{incr_8}} \cdot 
	 \transaction{t4}{\aTxnID[4] :\CommitTxnInv{\aTxnID[4]}{9}} \\
    \end{align*}
    \end{subfigure}

    \begin{subfigure}{\textwidth}
        \vspace{-8pt}
    \caption{Dependency graph.}
    \label{fig:ex-dependency-graph}
    \begin{tikzpicture}
    \tikzset{eventA/.style={draw, circle, minimum size=4pt, inner sep=1pt, fill}}
     \node[] (start) {};
     \node[eventA, right = 1.0cm of start] (e0) {  };
     \node[eventA, above right = 0.5cm and 2.0cm of e0] (e1) {};
     \node[eventA, below right = 0.5cm and 2.0cm of e0] (e2) {};
     \node[eventA, right = 3.5cm of e1] (e3) {};
     \node[eventA, right = 3.5cm of e2] (e4) {};  
     \node[above right = 0.5cm and 2.0cm of e4] (e5) {};
     \draw[->] (start) -- (e0);
     \draw[->] (e0) -- (e1);
     \draw[->] (e0) -- (e2);
     \draw[->] (e1) -- (e3);
     \draw[->] (e2) -- (e4);
     \draw[->] (e4) -- (e5);
     \draw[->] (e3) -- (e5);
     \draw[->] (e1) -- (e4);
     \node[text width=3.1cm, align=right, below left = 0.05cm of e0, t0] {\footnotesize $\TxnDescriptorInv{\aTxnID[0]}{0}{\_}{\{a\}}{\aEffectBuf_0}{1}$  \\ with $\aEffectBuf_0 = \{a \mapsto assign_0 \}$};
     \node[text width=3cm, align=right, above = 0.1cm of e1, t1] {\footnotesize $\TxnDescriptorInv{\aTxnID[1]}{2}{\_}{\{a\}}{\aEffectBuf_1}{4}$  \\ with $\aEffectBuf_1 = \{a \mapsto incr_1 \}$};
     \node[text width=3cm, align=right, above = 0.1cm of e3, t3] {\footnotesize $\TxnDescriptorInv{\aTxnID[3]}{5}{\_}{\{a\}}{\aEffectBuf_3}{8}$  \\ with $\aEffectBuf_3 = \{a \mapsto incr_4 \}$};
     \node[text width=3cm, align=right, below = 0.1cm of e2, t2] {\footnotesize $\TxnDescriptorInv{\aTxnID[2]}{2}{\_}{\{a\}}{\aEffectBuf_2}{6}$  \\ with $\aEffectBuf_2 = \{a \mapsto incr_2 \}$};
     \node[text width=3cm, align=right, below = 0.1cm of e4, t4] {\footnotesize $\TxnDescriptorInv{\aTxnID[4]}{7}{\_}{\{a\}}{\aEffectBuf_4}{9}$  \\ with $\aEffectBuf_4 = \{a \mapsto incr_8 \}$};
     \node[right = 0.1cm of e5] {$\LookupInv{\aStore}{a}{11} = 15$};
    \end{tikzpicture}
    \end{subfigure}

    \begin{subfigure}{0.3\textwidth}
        \vspace{-4pt}
    \caption{Map store for key $a$ at $t=11$.}
    \vspace{-4pt}
    \label{fig:ex-map}    
    \[
        \aMapStoreElem{a}{t} = 
        \begin{cases}
            \transaction{t0}{(0, assign_0)} & \text{for }\transaction{t0}{t = 1} \\
            \transaction{t1}{(2, assign_1)} & \text{for }\transaction{t1}{t = 4} \\
            \transaction{t2}{(2, assign_2)}  & \text{for }\transaction{t2}{t = 6} \\
            \transaction{t3}{(5, assign_5)}  & \text{for }\transaction{t3}{t = 8} \\
            \transaction{t4}{(7, assign_{11})}  & \text{for }\transaction{t4}{t = 9} \\
            \bottom & \text{otherwise}
        \end{cases}
    \]
    \end{subfigure}
    \begin{subfigure}{0.65\textwidth}
\vspace{-4pt}
     \caption{ Journal store at $t=11$.}    
     \vspace{-4pt}
    \label{fig:ex-journal}

%\[\begin{array}[t]{r@{}l}
%    \aStore = 
%    & \ \BeginTxnRecInv{\aTxnID[0]}{0} \append \UpdateRecInv[{\aTxnID[0]}]{a}{assign_0} \append \CommitTxnRecInv{\aTxnID[0]}{0}{1} \\
%    & \ \append\ \BeginTxnRecInv{\aTxnID[1]}{2} \append \UpdateRecInv[{\aTxnID[1]}]{a}{incr_1} \append \BeginTxnRecInv{\aTxnID[2]}{2} \\
%    & \ \append \UpdateRecInv[{\aTxnID[2]}]{a}{assign_2} \append \CommitTxnRecInv{\aTxnID[1]}{2}{4} \\
%    & \ \append\ \BeginTxnRecInv{\aTxnID[3]}{5} \append \CommitTxnRecInv{\aTxnID[2]}{2}{5} \append \UpdateRecInv[{\aTxnID[3]}]{a}{incr_4} \\
%    & \ \append\ \BeginTxnRecInv{\aTxnID[4]}{9} \append \CommitTxnRecInv{\aTxnID[3]}{5}{8} \append \UpdateRecInv[{\aTxnID[4]}]{a}{incr_8} \\
%    & \ \append\ \CommitTxnRecInv{\aTxnID[4]}{7}{9} \\
%  \end{array}\]
  
    \[\begin{array}[t]{r@{}l}
    \aStore = 
    & \ \transaction{t0}{\BeginTxnRecInv{\aTxnID[0]}{0}} \append 
    	\transaction{t0}{\UpdateRecInv[{\aTxnID[0]}]{a}{assign_0}} \append 
	\transaction{t0}{\CommitTxnRecInv{\aTxnID[0]}{0}{1}} \\
    & \ \append\ \transaction{t1}{\BeginTxnRecInv{\aTxnID[1]}{2}} \append 
    	\transaction{t1}{\UpdateRecInv[{\aTxnID[1]}]{a}{incr_1}} \append 
	\transaction{t2}{\BeginTxnRecInv{\aTxnID[2]}{2}} \\
    & \ \append \transaction{t2}{\UpdateRecInv[{\aTxnID[2]}]{a}{incr_2}} \append
     	\transaction{t1}{\CommitTxnRecInv{\aTxnID[1]}{2}{4}} \\
    & \ \append\ \transaction{t3}{\BeginTxnRecInv{\aTxnID[3]}{5}} \append 
    	\transaction{t2}{\CommitTxnRecInv{\aTxnID[2]}{2}{5}} \append 
	\transaction{t3}{\UpdateRecInv[{\aTxnID[3]}]{a}{incr_4}} \\
    & \ \append\ \transaction{t4}{\BeginTxnRecInv{\aTxnID[4]}{9}} \append 
    	\transaction{t3}{\CommitTxnRecInv{\aTxnID[3]}{5}{8}} \append 
	\transaction{t4}{\UpdateRecInv[{\aTxnID[4]}]{a}{incr_8}} \\
    & \ \append\ \transaction{t4}{\CommitTxnRecInv{\aTxnID[4]}{7}{9}} \\
  \end{array}\]
    \end{subfigure}
\end{minipage}
    \vspace{-4pt}
    \caption{Example of execution trace. The history (a) shows the order in which the transactional operations are executed. The dependency graph (b) visualizes the partial order of the transactions.
    The map (c) and journal (d) show the different stores after the history executed.}
    
  \label{fig:example-dependency-graph}
\end{figure*}

For interested readers, we provide a small-step operation semantics of 
transactions in the Appendix in Figure~\ref{fig:transaction-semantics}. %, using the standard notation.
Figure~\ref{fig:example-dependency-graph} illustrates these semantics with an example. The history (Figure~\ref{fig:ex-history}) shows a sequence of (atomic) transactional steps; the steps for concurrently executed transactions, like $\aTxnID[1]$ and $\aTxnID[2]$, are interleaved.
For simplicity, the example updates a single key $a$ of integer type.
Figure~\ref{fig:ex-dependency-graph} visualizes the transactions in a dependency graph.

\section{Store Models}
\label{sec:stores}

This section discusses two basic variants implementing the general store API\@.
% : the map-based and the journal-based store.
We aim to model their most essential, primitive properties, abstracting away as much complexity as possible.

\begin{figure*}[tp]
\begin{minipage}{0.9\hsize}\centering \small
  \[
  \sigma \in ( \TKey \times \TTimestamp \rightarrow \TTimestamp \times \TAssign)
\]
\[
  \begin{array}{rcl}
    \DoBeginInv{\aStore}{\_}{\_} &=& \aStore \\
    \LookupInvDefault 
      &=& \Merge \left(\max_{\visMS}( \{ \aMapStoreElem{\aKey}{\aVT}  \suchthat  \aVT < \aTS \})\right) \\
% &&\\
    \DoUpdateInv{\aStore}{\_}{\_}{\_}{\_}    & = & \aStore
    \\
    \DoCommitInv{\aStore}{\_}{\aST}{\_}{\aWriteSet}{\aEffectBuf}{\aCT}{[}\aKey,t{]}
    & = & 
      \begin{cases}
        (\aST, \aEffectBuf[\aKey])
            & \text{if}\ \aKey \in \aWriteSet \land t = \aCT
        \\
         \aMapStoreElem{\aKey}{t} & \text{otherwise}
       \end{cases}
  \end{array}
\]
\begin{flushleft}
%   where $  \aMapStoreElem{\_}{v_{1}} \visMS \aMapStoreElem{\_}{v_{2}}
%   \equaldef
%   v_{1} < \aMapStoreElem{\_}{v_{2}}.d$, and 
% $\max_{\visMS}$ is the set of highest elements in order ${\visMS}$.

\todo[Marc]{Prove that \DoCommit{} does not overwrite an existing
entry.}
\todo[AB]{This is only guaranteed if ct is different from any t
  in the key-timestamp pair; to prove it we need the transaction rules.}
\end{flushleft}
\end{minipage}
\caption{Semantics of map-based store.}
\vspace{-4pt}
  \label{fig:map-store-semantics}
\end{figure*}

\subsection{Map-based store semantics}
\label{sec:map-semantics}

The \emph{map-based store} models a classic versioned key-value store as a random-access map, located either in memory or on disk.
It is restricted to contain only values, which (in our model) are
represented as assignment effects.
Versions of a key are distinguished by their 
version timestamps.
Such a store maps a \emph{(key, version timestamp)} pair, to an \emph{(assignment effect, dependency timestamp)} pair.

Figure~\ref{fig:map-store-semantics} summarises its semantics.
A map store defers its updates to commit time, and committing atomically 
copies the transaction's \emph{effect buffer} into a new version of the corresponding keys.
\Lookup{} searches for the most recent assign effect directly from the map;
both \DoBegin{} and \DoUpdate{} are no-ops.
Figure~\ref{fig:ex-map} illustrates the contents of a map store, after
the history in Figure~\ref{fig:ex-history}.

In more detail, mapping $\aMapStoreElem{\aKey}{\aVT} = (d, \aEffect)$
associates a \emph{versioned key} $(\aKey, \aVT)$ with a \emph{dependent
  effect} $(d, \aEffect)$.
Here, $\aVT$ is a version timestamp, \aEffect is an assignment (a map store
does not support non-assignment effects) associated with
\emph{metadata} $d$, called dependence timestamp, where $d \le \aVT$.
Versions are ordered by their OTSPs $(d,\aVT)$, i.e.,
\[
  \aMapStoreElem{\aKey}{\aVT[1]} \visMS \aMapStoreElem{\aKey}{\aVT[2]}
  \equaldef
  \aVT[1] < \aMapStoreElem{\aKey}{\aVT[2]}.d
\]

%We assume that every transaction history forms a proper sequence.

When a transaction commits, method \DoCommit{} of a map store eagerly creates a new version for each key modified by the transaction.
The version identifier is the transaction's {commit timestamp}
\aCT, and it is associated with metadata \aST{}, the transaction's
\emph{snapshot timestamp}.
It returns a store unchanged except for the new versions.%
\footnote{
  A simpler specification might copy all keys, not just the dirty ones.
  However, this is not well-defined when the space of keys is unbounded
  (e.g., if keys are arbitrary strings).
}

The versions of \aKey{} visible from the current transaction are the set 
$V = \{ \aMapStoreElem{\aKey}{\aVT} \suchthat \aVT < \aTS \}$.
Since a map contains only assignments, \LookupInvDefault{} can omit all but the most recent one in this set in visibility order, noted $\max_{\visMS}(V)$.
% The versions of interest are those with timestamp $\aVT < \aTS$, i.e., the set 
% $\{ \aMapStoreElem{\aKey}{\aVT} \suchthat \aVT < \aTS \}$.
% Since a map contains only assignments, we can omit all but the most
% recent ones, and consider only $\max_{\visMS}(\ \{
% \aMapStoreElem{\aKey}{\aVT} \suchthat \aVT < \aTS \}\ )$, where
% $\max_{<}(S)$ denotes the set of maximal elements of set $S$ ordered by
% $<$.
To determine the returned value, any concurrent effects are merged (as
explained in Section~\ref{sec:system-model}), and the resulting
assignment effect is then applied to $\bottom$ to obtain a value.

The map store defers updates to commit time; therefore \DoUpdate leaves the store unchanged.

In practice, many existing database backends contain an in-memory map
store, for simplicity and fast reads.
To persist a map store, it suffices to write it to disk periodically;
however such a large write can be slow and is not natively crash-atomic.

\begin{figure*}[tp]
\begin{minipage}{0.9\hsize}\centering \small
  \begin{align*}
    \sigma \in \{\seq{e_1, e_2, \ldots} \suchthat e_i \in (
      &\{\BeginTxnRecInvDefault{} \suchthat \aTxnID{} \in \TTxnID, \aST \in \TTimestamp\}  \\ 
      & \cup \{\UpdateRecInv[\aTxnID]{\aKey}{\aEffect} \suchthat \aTxnID{} \in \TTxnID, k \in \TKey, \aEffect \in \TEffect\} \\ 
      & \cup \{\CommitTxnRecInv{\aTxnID}{\aST}{\aCT} \suchthat \aTxnID{} \in \TTxnID, \aST, \aCT \in \TTimestamp\})\}
\end{align*}

  \begin{flushleft}
    % A journal is a sequence of records; we note $<$ their sequential order.
    % There is also a partial visibility order $\visJS$
    % \todo[Marc]{Define $\visJS$ formally}.
    % The journal can be read in $<$ order, left to right, because
    % $\CommitTxnRecInv{\aTxnID[1]}{\aST[1]}{\aCT[1]} \visJS
    % \BeginTxnRecInv{\aTxnID[2]}{\aST[2]} \implies
    % \CommitTxnRecInv{\aTxnID[1]}{\aST[1]}{\aCT[1]} <
    % \BeginTxnRecInv{\aTxnID[2]}{\aST[2]}$.
    % ~\\[2ex]
    Note $\max_{\visJS}(\aRecord)$ the immediate predecessor(s) of record
    \aRecord{} in $\visJS$ order (\BeginTxnRec{} may have any number of mmediate
    predecessors; \UpdateRec{} and \CommitTxnRec{} have
    exactly one immediate predecessor).
    The \poststate{} function computes the state of a key
    \aKey{} after record \aRecord{} takes effect, as follows:
  \end{flushleft}
  ~\\  
  \begin{array}[t]{rcl}
    \poststateInvDefault & = &
      \begin{cases}
        \ \MergeSetInv{
                \poststateInv{\aRecord'}{\aKey}
                  \suchthat \aRecord' \in \max_{\visJS}{\aRecord{}}
        }
        & \text{if } \aRecord = \BeginTxnRecInv{\_}{\_} \\
        \ \poststateInv{\max_{\visJS}{\aRecord{}}}{\aKey{}}
        & \text{if } \aRecord{} = \UpdateRecInv[\_]{\aKey'}{\_}
                     \land \aKey{} \neq{} \aKey' \\
        \ \poststateInv{\max_{\visJS}{\aRecord{}}}{\aKey{}} \apply{} \aEffect{}
        & \text{if } \aRecord{} = \UpdateRecInv[\_]{\aKey}{\aEffect{}} \\
        \ \poststateInv{\max_{\visJS}{\aRecord{}}}{\aKey{}}
        & \text{if } \aRecord{} = \CommitTxnRecInv{\_}{\_}{\_} \\
      \end{cases}
  \end{array}
~\\
  \begin{flushleft}
    The journal operations are defined as follows (where $\append$
    represents the append operation):
  \end{flushleft}
 \[
 \begin{array}{rcl}   
       \DoBeginInvDefault & = &
          \aStore\ \append\ \BeginTxnRecInvDefault \\
      \LookupInv{\aStore}{\aKey}{\aTS} & = &
          \poststateInvDefault{} \text{ where } \aRecord = \BeginTxnRecInv{\_{}}{\aTS{}}
\\      
       \DoUpdateInvDefault{} & = &
          \aStore\ \append\ \UpdateRecInv[\aTxnID]{\aKey}{\aEffect}
   \\
      \DoCommitInv{\aStore}{\aTxnID}{\_}{\_}{\_}{\_}{\aCT}
                       & = & \aStore\ \append\ 
                             \CommitTxnRecInv{\aTxnID}{!!nothing!!}{\aCT}
   \\
  \end{array}
  \]
\end{minipage}
\caption{Semantics of journal-based store.}
  \label{fig:journal-store-semantics}
\end{figure*}

\subsection{Journal-based store semantics}
\label{sec:journal-semantics}

An alternative store variant is the journal-based store, which logs its updates
incrementally to a sequential file.%
\footnote{
  This describes a \emph{redo log}; an ``undo log'' would store inverse
  effects instead.
}
This design is optimised for fast disk writes, and has good
crash-tolerance properties.
It is also friendly to non-assignment effects.
However, to lookup the value of a key can be slow, as its semantics is
to read the journal and applies effects sequentially.

Figure~\ref{fig:journal-store-semantics} gives the formal semantics of a
journal store.
A journal store is a finite sequence $\aStore = \seq{e_{1}, e_{2}, \ldots{}}$
of records of type BeginTxnRec{}, \UpdateRec{} and \CommitTxnRec{},
initially empty.
Function \DoBegin{} appends a record with transaction identifier
\aTxnID{} and snapshot timestamp \aST{}.
\DoUpdate{} appends an \UpdateRec{} record that contains transaction
identifier \aTxnID{}, key \aKey{}, and effect \aEffect{}.
Similarly, \DoCommit{} appends a \CommitTxnRec{} record containing the
transaction identifier \aTxnID{}, snapshot timestamp \aST{} and commit
timestamp \aCT{}.

The real action is in \LookupInvDefault{}, which accumulates the effects
to key \aKey{} that committed strictly before \aTS.
To formalise the procedure is somewhat complex.
\begin{compactitem}
\item
  Procedure \poststateInvDefault{} computes the state of key \aKey{}
  after a record \aRecord{} in journal \aStore{} takes effect.
  Records take effect in \visJS{} order.
\item
  In \visJS{} order, a record of type \BeginTxnRec{} has any number of
  immediate predecessors; other types of records have a single one.
\item
  The poststate of an update record with key \aKey{} and effect
  \aEffect{} is computed by taking the poststate of its
  immediately-preceding record, and applying \aEffect{}.
\item
  The poststate of a \BeginTxnRec{} is the \Merge{} of poststate of its
  immediate predecessors.
\item
  Otherwise, the poststate is the same as that of the immediate
  predecessor; i.e., updates for other keys are ignored, as well as commit
  records.
\end{compactitem}
Note that a poststate can be computed in a single left-to-right pass
over the journal, because a \CommitTxnRec{} record always appears before
the \BeginTxnRec{} of a transaction that depends on it.
Note also that records are single-assigned and that \poststate{} is a
function; therefore a practical implementation may use a cache.

Figure~\ref{fig:ex-journal} illustrates the contents of a journal store
after execution of the history in Figure~\ref{fig:ex-history}.
To obtain the value for $\LookupInv{\aStore}{x}{11}$, we calculate
recursively the \poststate{}s for $\aTxnID[3]$ and $\aTxnID[4]$ and
merge the results:
\begin{small}
\begin{align*}  
  \LookupInv{\aStore}{a}{11} &= \Assign[{0}] \apply \MergeSetInv{ \{\Incr[5], \Incr[{11}]\} } \\
  & = \Assign[{15}]
\end{align*}
\end{small}

As explained earlier (Section~\ref{sec:system-model}) the implementation of effects must ensure that the \Incr[1] from $\aTxnID[1]$ executes once only in the merged value, even though the history contains two paths to \Lookup{}.

\section{Formal model in Coq}
\label{sec:coq}
So far, we have formalized a major part of the definitions presented in Section~\ref{sec:system-model}, \ref{sec:stores} and \ref{sec:transactions} in the proof assistant Coq with the goal to formally verify their correctness.
Our Coq codebase comprises currently around 2k LOC, without using any external libraries other than the standard library.
%This mostly includes definitions presented in this paper and some properties about them, but the majority of proofs are still in progress.

There are several reasons for choosing to use a proof tool such as Coq.
A positive aspect about the Coq formalization is the high level of abstraction.
We can define constructs with desired properties without the need to provide a specific implementation.
This is in contrast to traditional programming languages, where interfaces or abstract classes can be defined, but it is usually not possible to restrict the behavior of implementations.
As an example of this, in our formalization we defined timestamps to be some arbitrary data type that comes with an ordering, as described in section~\ref{sec:system-model}, and for which equality is decidable.
No assumption about the implementation is made, and refining the specification to specific instances can be done independently.

However, the main benefit of reasoning about our system in a formal context is the required level of detail and precision, which is typically not attained by pen-and-paper proofs.
It not only makes spotting mistakes easier and earlier, but it also forced us to pay attention to specific corner cases.
This already proved to be useful during this initial formalization phase.
For example, in earlier versions the journal-based store explicitly maintained both a dependency and commit timestamp, while the map-based store did this implicitly.
This oversight lead to an inconsistency when merging concurrent effects, since the map-based store carries too little information.
Only when trying to formalize the map-based store, we noticed this mismatch.

% \todo[AB]{This section is a bit too specific maybe.}
% Naturally, with the use of a formal system such as Coq, we also face challenges.
% Currently, our biggest inconvenience in the formalization comes from the fact that we have to work with finite maps.
% In Coq's standard library these are defined using a custom equality (called a setoid equality), different from the normal equality of Coq.
% This results in the fact that we have to carry the proofs of equivalences between maps into virtually every other proof.
% While not being extremely difficult to work with, as Coq has relatively advanced setoid rewrite systems, it is sometimes a hassle to work with.
%Since most of the formalization at the moment consists of definitions, we will probably encounter further challenges while progressing in proving properties.

% The choice of using Coq for our formalization was mainly guided by the proof assistant's maturity.
% With a big and active community, as well as advanced techniques for automating proofs, it is one of the most advanced proof assistants.
% By now Coq is also used in different fields of computer science not traditionally associated with formal methods or directly connected to the typical uses in programming language theory or pure mathematics.
% These aspects, as well as our already existing familiarity with the tool, made us decide to formalize our model using Coq.

% We are currently in the middle of the formalization phase in Coq, and it already helped in discovering errors or inconsistencies in our formal model.
While it is arguably more effort to formalize everything in a proof assistant rather than using pen-and-paper definitions and proofs, we believe that the benefits of obtaining a verified design outweigh the costs.

\section{Discussion and Outlook}
\label{sec:discussion-outlook}

Formal methods have been successfully employed for proving (distributed) systems correct~\cite{DBLP:journals/toplas/Lamport94, DBLP:conf/pldi/WilcoxWPTWEA15, DBLP:journals/cacm/HawblitzelHKLPR17,Chang23}.
The focus of these approaches has been the verification of safety and liveness properties for different types of distributed systems and their implementations. 
The work presented in~\cite{Chang23} is closest to our approach, as it proves the correctness of a transaction library.
However, their work targets the correctness verification of the specific library and all its sophisticated optimisations,
while we aim to take a compositional approach to proving a generic database backend.

%% Carla: There are work much closer related that we don't mention...

%In\~cite{Haller18}, the authors propose a programming model where data remains stationary and computations are mobile. 
%While their formalisations share some similarities with ours, the storage models explored are orthogonal.

Typically, developers need to provide a high-level specification that is then refined in one or more steps, while the corresponding proofs are correspondingly refined.
For example, Verdi~\cite{DBLP:conf/pldi/WilcoxWPTWEA15} is a framework to implement and specify systems under different network  semantics in Coq; starting from an idealized system, proof obligations are then transformed for more and more complex fault models. 
Finally, an implementation can then be generated from the specification. 
Our approach differs in two major aspects.
The focus of our work is the correct design and implementation of a central system component, not a distributed protocol. This component is typically simplified in the before mentioned verification frameworks.
However, a correct and efficient implementation is essential in any (distributed) datastore.
Further, instead of refinement, we propose a compositional approach to construct more and more complex implementations. 
This helps system designers to select and incrementally add features such as caching, write-ahead logging, or checkpointing.
Reducing these features to their essence helps us extracting their actual (and not incidental) requirements and to re-purpose metadata in different contexts.

Tools exist to compile a Coq specification to executable code
\cite{DBLP:journals/jip/TanakaAG18,DBLP:conf/cie/Letouzey08, DBLP:conf/pldi/WilcoxWPTWEA15}.
We are not taking this path for pragmatic reasons: it is too far
from the ordinary programmer's experience and as of today still requires extensive
manual intervention.
Instead, we manually transcribe the specification to
Java, verbatim, resisting the temptation to optimise. We will check through 
testing that the implementation behaves like the specification, and that
variants that were proved equivalent do have the same runtime behaviour.

% \begin{acks}t
%   To Robert, for the bagels and explaining CMYK and color spaces.
% \end{acks}

\balance
\bibliographystyle{ACM-Reference-Format}
\bibliography{predef,shapiro-bib-ext,shapiro-bib-perso,local, bibliography}

%\pagebreak
\appendix

\section{Formal semantics of transactions}
\label{app:transaction-semantics}

Figure \ref{fig:transaction-semantics} shows the transition rules for the small-step operational semantics of a transaction system build on a store.
The specification is fully formal and unambiguous: we find it invaluable
to reason about the system, and it is easily translated to the language
of a proof tool such as Coq.
Most interestingly, it can be read as pseudocode, as we explain now.

The semantics are written as a set of rules.
Each rule represents an indivisible state transition; i.e., there are no
intermediate states from a semantic perspective, and any intermediate
states in the implementation must not be observable.

The system state is represented as a tuple $\aStateInvDefault$ consisting of a
store, its field, and the sets of aborted, committed, and running
transactions' descriptors.

A rule consists of a set of \emph{premises} above a long horizontal
line, and a \emph{conclusion} below.
A premise is a logical predicate referring to state variables.
A variable without a prime mark refers to before the state before the
transition (pre-state); a primed variable refers to state after the
transition (post-state).
Thus a premise that uses only non-primed variables is a pre-condition on
the pre-state; if it contains a primed variable, it is a post-condition
that constrains the post-state.

If the premises are satisfied, the state-change transition described by
the conclusion can take place.
A label on the transition arrow under the line represents a client API
call.
Thus, a rule can be seen as terse pseudocode for the computation to be
carried out by the API.

\subsection{Example}

To explain the syntax, consider for example rule \StartTxnRule{}.
The conclusion describes the transition made by API command \BeginTxnInvDefault{}
from pre-state \aStateInvDefault{} on the left
of the arrow $\xrightarrow[\aTxnID]{\BeginTxnInvDefault}$, to post-state
$\aStateInv[\aStore']{\aField[\aStore]}{\abortedSet}{\committedSet}{\runningSet'}$ on the right.
Note that in the right-hand side of this conclusion, only \runningSet{} is
primed, indicating that the other elements of the state do not change.

% The rules are also parameterised for consistency model, via
% preconditions \CMbegin{} and \cmcommit.
% In our baseline TCC consistency model, they are empty (i.e., true).
% In future work, they will be adapted to support stronger consistency
% models.

\subsection{Parameters}

The rules describe a transaction system, which is a tuple \aStateInvDefault{}
consisting of a store \aStore{}, its associated field \aField{},
and sets of transaction descriptors
{\abortedSet}, {\committedSet}, and {\runningSet}, which keep track of aborted,
committed and running (ongoing) transactions respectively.
A transaction descriptor is a tuple \TxnDescriptorDefault{} of transaction identifier \aTxnID,
its \emph{snapshot timestamp} \aST, its \emph{read set} \aReadSet,
its \emph{write set} \aWriteSet, its \emph{effect buffer} \aEffectBuf, and
its \emph{commit timestamp} \aCT.%
\footnote{
  To simplify notation, we may write $\aTxnID.\aST$ or
  $\aTxnID.\aCT$, for instance, for the corresponding elements of
  the descriptor whose transaction identifier is \aTxnID{}.
  % We may write $\Pi_{\aST}(\aTxnSet{})$ for the projection of the set
  % of transaction descriptors \aTxnSet{} along the \aST{} dimension.
}

The two timestamps define visibility between transactions, as defined
previously (Section~\ref{sec:system-model}).
Initially, after rule \StartTxnRule{}, the sets and the effect buffer are empty
and the commit timestamp is invalid.
For each key that is accessed, rule \InitKeyRule{} initialises the
buffer, and rule \UpdateRule{} updates it.
Computation of the actual commit timestamp may be deferred to the
\CommitRule{} rule.

The semantic rules are parameterised by commands \Lookup{}, \DoUpdate{}, and
\DoCommit{}, specified in Figure~\ref{fig:store-interface}.
These commands are specialised for each specific store variant: the map-based variant in
Section~\ref{sec:map-semantics}, the journal-based variant in
Section~\ref{sec:journal-semantics}.
% The rules are also parameterised for consistency model, via
% preconditions \CMbegin{} and \cmcommit.
% In our baseline TCC consistency model, they are empty (i.e., true).
% In future work, they will be adapted to support stronger consistency
% models.

\subsection{Transaction begin}

We now consider each rule in turn.

\StartTxnRule{} describes how API $\BeginTxnInv$ begins a new
transaction with snapshot timestamp \aST{}.
The snapshot of the new transaction is timestamped by \aST, passed
as an argument; remember that a snapshot includes all transactions that
committed with a strictly lesser commit timestamp.

The first premise chooses a fresh transaction identifier
\aTxnID{}.
The last premise ensures that the appropriate transaction descriptor
is in the post-state set of running transactions.

% The \emph{visibility} premise $\forall t \in \Pi_{\aCT}(\runningSet): t
% \not< \aST$ states that the new transaction may not depend on some
% transaction that is not yet terminated.
% Suppose this premise was not present: then this transaction might attempt
% to read a value that has not yet been written or will be aborted.
% We discuss the synchronisation required for enforcing it in
% Section~\ref{sec:enforce-visibility}.

As the transition is labeled by \aTxnID{}, multiple instances of
\StartTxnRule{} are mutually independent and might execute in parallel,
as long as each such transition appears atomic.

\subsection{Reads and writes}

Reading or updating operate on the transaction's effect buffer
\aEffectBuf{}, which must contain the relevant key.

Rule \InitKeyRule{} specifies a \emph{buffer miss}, which initialises
the buffer for some key \aKey{}.
As it does not have an API label, it can be called arbitrarily.
It modifies only the current transaction's descriptor.
Its first premise takes the descriptor of the current transaction
\aTxnID{} from the set of running transactions \runningSet{}.
The second one checks that \aKey{} is not already in the read set,
ensuring that the effect buffer is initialised once per key.
The third reads the store by using \Lookup{} (specific to a store
variant).
Next, a premise updates the read set, and another initialises the effect
buffer with the return value of \Lookup{}.
The final premise puts the transaction descriptor, containing the
updated read set and effect buffer, back into the descriptor set of
running transactions.

In Rule \ReadRule{}, API $\ReadInv{\aKey}$ returns a value from the
effect buffer.
It does not modify the store.
The first premise is as above.
The second one requires that the key is in the read set, thus ensuring
that \InitKeyRule{} has been applied.
The next two premises extract \aKey{}'s mapping from the effect buffer
and compute the corresponding return value.

Note the clause \(\aEffect \in \TAssign\).
It requires that, previously to reading, the application has initialised
the store with an assignment to \aKey{} (possibly followed by other
effects; such a sequence resolves to an assignment, by associativity, as
explained earlier).
Otherwise, \Lookup{} would return either \bottom{} (if the key has not
been initialised at all) or a non-assignment (if the application has
stored only non-assignments).
We leave the burden of initialisation to the application to simplify the
semantics;
% especially when considering store
% composition (Section~\ref{sec:store-composition}).Logically, 
logically, it's an axiom.

In Rule \UpdateRule{}, API call \UpdateInv{\aTxnID}{\aKey}{\aEffect} applies
effect \aEffect{} to key \aKey{}.
It updates both the store and the transaction descriptor.
The first two clauses are similar to \ReadRule{}, and similarly require
a buffer miss if the key has not been used before (avoiding blind
writes).
It updates the effect buffer, ensuring that the transaction will read its
own writes, and puts the key in the write set.
It calls the variant-specific command \DoUpdate{}, discussed later in
the context of each variant.

\subsection{Transaction termination}

A transaction terminates, either by aborting without changing the store,
or by committing, which applies its effects atomically to the store.
% Note that the \AbortRule{} transition is not labelled, i.e., a
% transaction may abort spontaneously.

Rule \AbortRule{} moves the current transaction's descriptor
from \runningSet{} to \abortedSet{}, marking it as aborted.
It does not make any other change.

API call $\CommitTxnInv{\aTxnID}{\aCT}$ takes a commit timestamp argument.
It is enabled by rule \CommitRule{}, which modifies the store, the
running set, and the committed set.
The first premise is as usual.
Commit timestamp \aCT{} must
satisfy the constraints stated in the next three premises:
\begin{inparablank}
\item
  it is unique (it does not appear in \committedSet{});
  % , by premise $\aCT \notin
  % \Pi_{\aCT}(\committedSet)$.
\item
  it is greater or equal to the snapshot;
\item
  the \NIct{} premise ensures that no
  already-committed or running transaction may depend on this one.
\end{inparablank}

The latter premise aims to protect against the case where
another transaction has read a value that this transaction has yet to
write, because the transactions commit in the wrong order.
To understand, consider the following anomalous example:
\begin{inparaenum}[\it (i)]
\item
  Transaction $\aTxnID[1]$ has commit timestamp 1;
\item
  Transaction $\aTxnID[2]$ starts with snapshot timestamp $2>1$; thus $\aTxnID[2]$
  reads the updates made by $\aTxnID[1]$;
\item
  however, $\aTxnID[1]$ is slow and its committed effects reach the store only
  after the read by $\aTxnID[2]$.
\end{inparaenum}
Clearly, this would be incorrect.
To avoid this issue, $\aTxnID[2]$ must not start until $\aTxnID[1]$ has finalised its
transition to committed.
This requires synchronisation between concurrent transactions.
% which will be discussed in Section~\ref{sec:enforce-RC}.

To avoid this issue, no still-running or committed transaction may read from
the current transaction, i.e., \[\NIctdef{}
  \land \aWriteSet \intersect \aTxn{}.\aReadSet \neq \emptyset\]
%.
This expression requires to keep track of the read-set of committed
transactions; to avoid this, we could check running transactions only,
using the slightly stronger expression: \[\NIctdef\ \land ( \aTxn{} \in
\runningSet \implies \aWriteSet \intersect \aTxn{}.\aReadSet \neq
\emptyset ) \]%.
For simplicity, we choose to use the even stronger premise
\[\NIct \equaldef \NIctdef{}\]%,
at the cost of aborting transactions
unnecessarily.

Operation \DoCommit{} (specific to a store
variant) provides the new state of the store; it should ensure that the
effects of the committed transaction become visible in the store,
labelled with the commit timestamp.
Finally, the transaction descriptor, now containing the commit
timestamp, is moved to the set of committed transactions.

\makeatletter
\def\moverlay{\mathpalette\mov@rlay}
\def\mov@rlay#1#2{\leavevmode\vtop{%
   \baselineskip\z@skip \lineskiplimit-\maxdimen
   \ialign{\hfil$\m@th#1##$\hfil\cr#2\crcr}}}
\newcommand{\charfusion}[3][\mathord]{
    #1{\ifx#1\mathop\vphantom{#2}\fi
        \mathpalette\mov@rlay{#2\cr#3}
      }
    \ifx#1\mathop\expandafter\displaylimits\fi}
\makeatother

\newcommand{\cupdot}{\charfusion[\mathbin]{\cup}{\cdot}}

\begin{figure*}[tp]
\begin{minipage}{0.9\hsize}\centering \small
    \[
    \inferrule%[\StartTxnRule]
    {
      \forall \aTxn \in \abortedSet \cupdot \committedSet \cupdot \runningSet,
          \aTxn.\aTxnID \neq \aTxnID{}
      % \aTxnID \notin (\abortedSet \cup \committedSet \cup \runningSet).\aTxnID
      \\
      \aStore' = \DoBeginInvDefault
      \\
      \runningSet' = \runningSet\ \cupdot\ \{ \TxnDescriptorInv{\aTxnID}{\aST}{\emptyset}{\emptyset}{\emptyset}{\_} \}
    }
    {
      \aStateInvDefault \xrightarrow[\aTxnID]{\BeginTxnInvDefault}
      \aStateInv[\aStore']{\aField[\aStore]}{\abortedSet}{\committedSet}{\runningSet'}
    }
    \quad \textsc{[\StartTxnRule\!\!]}
  \]

  \[
    \inferrule%[\InitKeyRule]
    {
      \runningSet = \runningSet'' \cupdot \{\TxnDescriptorInv{\aTxnID}{\aST}{\aReadSet}{\aWriteSet}{\aEffectBuf}{\_}\}
      \\
      \aKey \notin {\aReadSet}
      \\
      \LookupInv{\aStore}{\aKey}{\aST} = \aEffect
      \\
      \aReadSet' = \aReadSet \cupdot \{ \aKey \}
      \\
      \aEffectBuf' = \aEffectBuf [\aKey \leftarrow \aEffect ]
      \\
      \runningSet' = \runningSet''
      \cupdot \{\TxnDescriptorInv{\aTxnID}{\aST}{\aReadSet'}{\aWriteSet}{\aEffectBuf'}{\_}\}
    }
    {
      \aStateInvDefault \xrightarrow[\aTxnID]{}
      \aStateInv{\aField[\aStore]}{\abortedSet}{\committedSet}{\runningSet'}
    }
    \quad \textsc{[\InitKeyRule\!\!]}
  \]

  \[
    \inferrule%[\ReadRule]
    {
      \runningSet = \runningSet'' \cupdot \{\TxnDescriptorInv{\aTxnID}{\aST}{\aReadSet}{\aWriteSet}{\aEffectBuf}{\_}\}
      \\
      \aKey \in {\aReadSet}
      \\
      \aEffectBuf[\aKey] = \aEffect  \in \TAssign
      \\
      \aValue = \aEffect(\bottom)
    }
    {
      \aStateInvDefault \xrightarrow[\aTxnID]{\ReadInv{\aKey} \rightarrow \aValue}
      \aStateInvDefault
    }
    \quad \textsc{[\ReadRule\!\!]}
  \]

  \[
    \inferrule%[\UpdateRule]
    {
      \runningSet = \runningSet''
        \cupdot \{\TxnDescriptorInv{\aTxnID}{\aST}{\aReadSet}{\aWriteSet}{\aEffectBuf}{\_}\}
      \\
        \aKey \in \aReadSet
      \\
        \aStore' = \DoUpdateInvDefault{}
        \\
        \aWriteSet' = \aWriteSet{} \union \{ \aKey \}
        \\
        \aEffectBuf' = \aEffectBuf{}[\aKey \leftarrow
                                      \aEffectBuf{} [\aKey] \apply \aEffect{} 
                                  ]
        \\
      \runningSet' = \runningSet''
        \cupdot \{\TxnDescriptorInv{\aTxnID}{\aST}{\aReadSet}{\aWriteSet'}{\aEffectBuf'}{\_}\}
    }
    {
      \aStateInvDefault \xrightarrow[\aTxnID]{\UpdateInv{\aTxnID}{\aKey}{\aEffect}}
      \aStateInv[\aStore']{\aField[\aStore]}{\abortedSet}{\committedSet}{\runningSet'}
    }
    \quad \textsc{[\UpdateRule\!\!]}
  \]

  \[
    \inferrule%[\AbortRule]
    {
      \runningSet = \runningSet'
      \cupdot \{\TxnDescriptorInv{\aTxnID}{\aST}{\aReadSet}{\aWriteSet}{\aEffectBuf}{\_}\}
      \\
      \abortedSet' = \abortedSet
      \cupdot \{\TxnDescriptorInv{\aTxnID}{\aST}{\aReadSet}{\aWriteSet}{\aEffectBuf}{\_}\}
    }
    {
      \aStateInvDefault \xrightarrow[\aTxnID]{\AbortTxnInv{}}
      \aStateInv{\aField[\aStore]}{\abortedSet'}{\committedSet}{\runningSet'}
    }
    \quad \textsc{[\AbortRule\!\!]}
  \]

  \[
    \inferrule%[\CommitRule]
    {
      \runningSet = \runningSet'
      \cupdot \{\TxnDescriptorInv{\aTxnID}{\aST}{\aReadSet}{\aWriteSet}{\aEffectBuf}{\_}\}
      \\
      \forall \aTxn \in \committedSet, \aTxn.\aCT \neq \aCT 
      % \aCT \notin \committedSet.\aCT
      \\
      \aST \le \aCT
      \\
      \NIct[\aCT] % non-inversion premise
      \\
      % \CCcommitInv[]{\aStore}{\committedSet}{\aCT}{\aReadSet}{\aWriteSet}{\aEffectBuf}{\committedSet}
      % \\
      \aStore' = \DoCommitInvDefault{}
      \\
      \committedSet' = \committedSet
      \cupdot \{\TxnDescriptorInv{\aTxnID}{\aST}{\aReadSet}{\aWriteSet}{\aEffectBuf}{\aCT}\}
    }
    {
      \aStateInvDefault \xrightarrow[\aTxnID]{\CommitTxnInv{\aTxnID}{\aCT}}
      \aStateInv[\aStore']{\aField[\aStore]}{\abortedSet}{\committedSet'}{\runningSet'}
    }
     \quad \textsc{[\CommitRule\!\!]}
  \]
\end{minipage}
\caption{Operational semantics of transactions. $\cupdot$ denotes disjoint set union.}
  \label{fig:transaction-semantics}
\end{figure*}

\end{document}